\documentclass[prl,twocolumn,superscriptaddress,showpacs,preprintnumbers,amsmath,amssymb]{revtex4}
\usepackage{graphicx}
\usepackage{dcolumn}
\usepackage{bm}
\usepackage{mathrsfs}
\usepackage{subfigure}
\usepackage{natbib}

\begin{document}

\title{Large Magnetoresistance Oscillations in Mesoscopic Superconductors Due to Current-Excited Moving Vortices}

\author{G. R. Berdiyorov}
\affiliation{Departement Fysica, Universiteit Antwerpen,
Groenenborgerlaan 171, B-2020 Antwerpen, Belgium}
\author{M. V. Milo\v{s}evi\'{c}}
\affiliation{Departement Fysica, Universiteit Antwerpen,
Groenenborgerlaan 171, B-2020 Antwerpen, Belgium}
\author{M. L. Latimer}
\affiliation{Materials Science Division, Argonne National Laboratory, Argonne, Illinois 60439, USA}
\affiliation{Department of Physics, Northern Illinois University, DeKalb, Illinois 60115, USA}
\author{Z. L. Xiao}
\email{xiao@anl.gov}
\affiliation{Materials Science Division, Argonne National Laboratory, Argonne, Illinois 60439, USA}
\affiliation{Department of Physics, Northern Illinois University, DeKalb, Illinois 60115, USA}
\author{W. K. Kwok}
\affiliation{Materials Science Division, Argonne National Laboratory, Argonne, Illinois 60439, USA}
\author{F. M. Peeters}
\email{francois.peeters@ua.ac.be}
\affiliation{Departement Fysica, Universiteit Antwerpen,
Groenenborgerlaan 171, B-2020 Antwerpen, Belgium}

\date{\today}

\begin{abstract}
We show in the case of a superconducting Nb ladder that a mesoscopic superconductor typically exhibits
magnetoresistance oscillations whose amplitude and temperature dependence are different from those stemming from the
Little-Parks effect. We demonstrate that these large resistance oscillations (as well as the monotonic background on
which they are superimposed) are due to {\it current-excited moving vortices}, where the applied current in competition
with the oscillating Meissner currents imposes/removes the barriers for vortex motion in increasing magnetic field. Due
to the ever present current in transport measurements, this effect should be considered in parallel with the
Little-Parks effect in low-$T_c$ samples, as well as with recently proposed thermal activation of dissipative
vortex-antivortex pairs in high-$T_c$ samples.
\end{abstract}


\maketitle

The circulation of electrons and Cooper pairs in doubly and multiply connected normal and superconducting materials in
the presence of a magnetic field produces respectively the Aharonov-Bohm \cite{Lee} and the Little-Parks (LP) effect
\cite{Little}. These effects are sensitive probes for the proposed incoherent Cooper pairing in the pseudogap
\cite{Stewart} and for insulating phases \cite{Carillo} competing with superconductivity in copper oxide and
conventional superconductors, for the relative contribution of individual bands in two-band superconductors \cite{Erin}
and for the interplay of superconductivity and magnetism in hybrid structures \cite{Samokhvalov}. They are expected to
shed light on the d-wave symmetry \cite{Barash} and the strip structures \cite{Berg} in high temperature
superconductors. Magnetoresistance oscillations have been widely observed in doubly and multiply connected mesoscopic
superconductors \cite{loop,loop1,loop2,NETt,NETe,NETr,Sochnikov,Kirtley} and have mostly been attributed to the LP
effect, which is related to the periodic suppression of the critical temperature $T_c$ due to fluxoid quantization
\cite{London,Deaver}. However, analysis of magnetoresistance oscillations in recent experiments on high-$T_c$
superconducting loops (in a specifically designed network), showed that neither the amplitude of the oscillations, nor
their temperature dependence could be explained by the LP effect \cite{Sochnikov}. In order to describe the
experimental data, the authors used the fluxoid dynamics model \cite{Kirtley}, according to which \textit{thermally
excited} vortices/antivortices move and interact with the magnetic field-induced persistent currents.

Here, we present results of numerical simulations and transport measurements, which reveal a new origin for the
magnetoresistance oscillations in mesoscopic superconductors, namely, the motion of current-excited vortices. These
vortices are not excited by thermal fluctuations in contrast to the case of Refs. \cite{Sochnikov,Kirtley}. Instead,
the applied current interacts with existing Meissner currents in the sample and tunes the barrier for vortex entry and
exit while simultaneously driving the vortices. In addition, the ubiquitous monotonic background of the
magnetoresistance curves is also found to be due to moving vortices. Since an external current is inherent to all
transport measurements, the proposed mechanism needs to be considered in the studies of singly (stripes), doubly
(loops) or multiply (ladders and wire networks) connected mesoscopic superconductors.

\begin{figure}[b]
\includegraphics[scale=0.3]{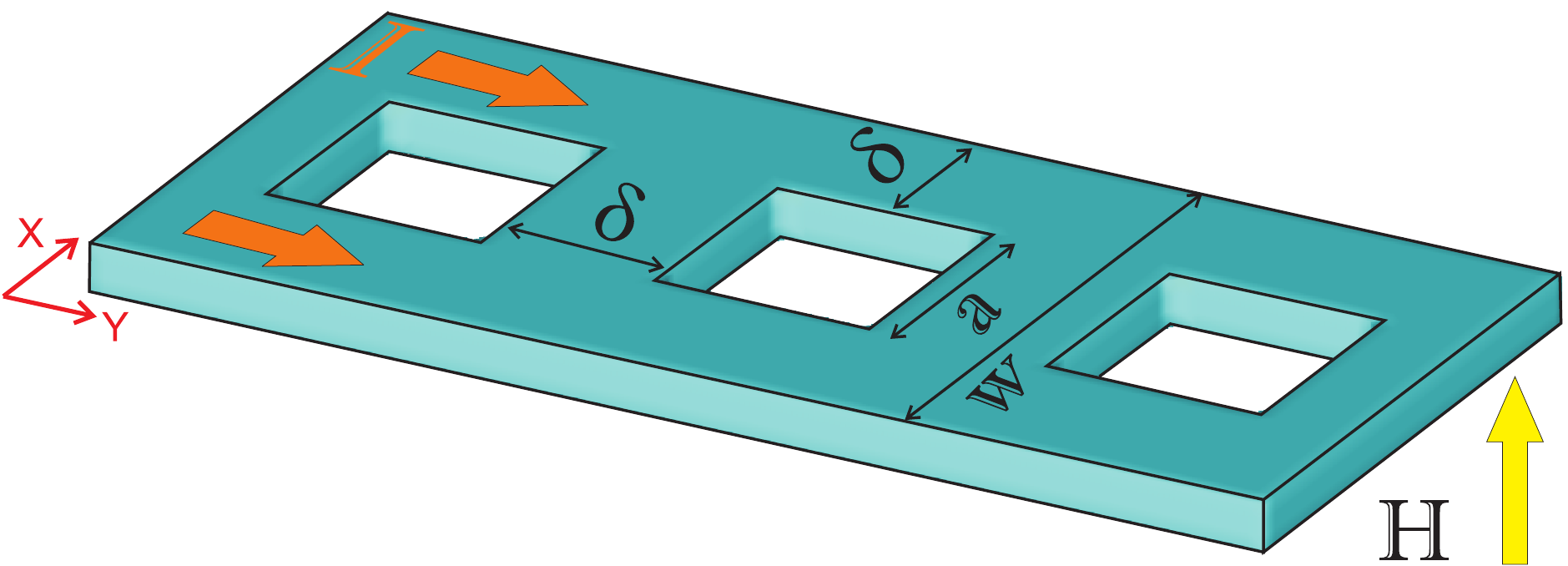}
\caption{\label{fig1}(color online) An oblique view of the system: a superconducting strip (of width $w=a+2\delta$, thickness $t\ll \lambda, \xi$, and periodically long in the $y$-direction) with a chain of holes (size $a$ and period $d=a+\delta$) under applied dc current $I$ and a perpendicular magnetic field $H$.}
\end{figure}

In what follows, we demonstrate our findings for a Nb ladder structure (i.e., a thin strip with a chain of holes, see
Fig. \ref{fig1}), which is a transition geometry from a finite ring structures to an infinite network \cite{ladder}. In
this geometry, the electrical contact issues on the resistance, known to occur in individual loops \cite{denis_heat},
are suppressed. The choice of material and geometry was made to maximize the larger characteristic length scales, to
avoid the formation of ``effective'' weak links, which can result in direct superconducting-insulator transitions
\cite{vinokur} and consequently large magnetoresistance oscillations in a broad range of temperatures and magnetic
fields \cite{baturina}. Finally, thermal fluctuations in Nb are far less important than in high-$T_c$ samples, which is
also relevant for the analysis of our findings in comparison with Ref. \cite{Sochnikov}.

\textit{Theory}. -- For the theoretical study of the fluxoid quantization and dynamics, we used the generalized
time-dependent Ginzburg-Landau (GL) equations \cite{Kramer}:

\begin{eqnarray}
\frac {u}{\sqrt{1+\gamma^2|\psi|^2}}\left(\frac{\partial}{\partial t}+\frac {\gamma ^2}{2}\frac {\partial
|\psi|^2}{\partial t}\right)\psi=(\bigtriangledown-i\mathbf{A})^2\psi\nonumber\\
+\left(1-T-|\psi|^2\right)\psi+\chi({\bf r},t),
\end{eqnarray}
\begin{eqnarray}
\frac{\partial {\bf A}}{\partial
t}=\textrm{Re}\left[\psi^*(-i\nabla-{\bf A})\psi\right]-\kappa^2
\textrm{rot rot} {\bf A},
\end{eqnarray}
with units: the coherence length $\xi(0)$ for the distance, $t_{GL}(0)$$=$$\pi \hbar \big/8k_BT_cu$ for time, $T_c$ for
temperature, and $\phi_0/2\pi\xi(0)$ for the vector potential ${\bf A}$. Order parameter $\psi$ is scaled to its value
for zero magnetic field and temperature, and the current density is in units of $j_0=\sigma_n\hbar/2et_{GL}(0)\xi(0)$.
The material parameters $\gamma=20$ and $u=5.79$ follow from microscopic theory \cite{Kramer}. $\chi({\bf r},t)$ is the
random fluctuating term \cite{kato}. To account for heating effects, we couple Eqs. (1-2) to the heat transfer equation
(see Ref. \cite{denis_heat} for details):
\begin{eqnarray}
\nu\frac{\partial T}{\partial t}=\zeta \nabla^2 T+\left(\frac{\partial
{\bf A}}{\partial t}\right)^2-\eta(T-T_0),
\end{eqnarray}
where $T_0$ is the bath temperature. Here we use $\nu$=0.03, $\zeta$=0.06 and $\eta=2\cdot10^{-4}$, corresponding to an
intermediate heat removal to the substrate \cite{denis_heat}. The above equations are solved self-consistently using
the semi-implicit Crank-Nicholson algorithm \cite{Winiecki} with periodic boundary conditions in the $y$-direction and
Neumann boundary conditions at all sample edges. The transport current is introduced via the boundary condition for the
vector potential, rot${\bf A}|_z(x$=$0,w)$=$H\pm H_I$, where $H_I$=$2\pi I/c$ is the magnetic field induced by the
current $I$.

\begin{figure}[t]
\includegraphics[width=\linewidth]{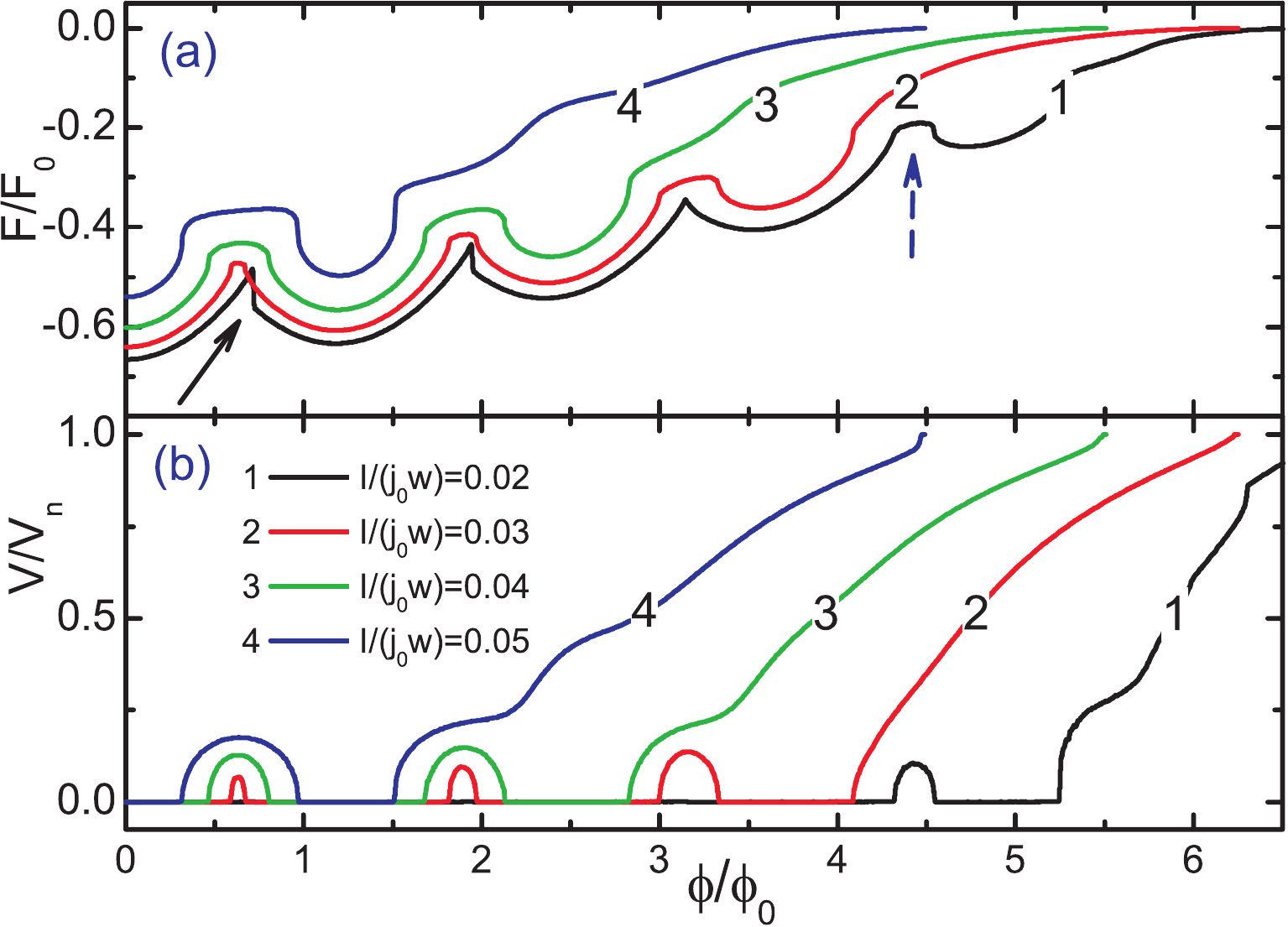}
\caption{\label{fig2}(color online) Time-averaged free energy  $F$ (a) (in units of $F_0=H_c^2V_0/8\pi$ with $V_0$ the
unit cell volume) and voltage $V$ (b) (in units of normal state voltage $V_n$) as a function of external flux $\phi$
(calculated over the unit cell area $S=w\times(a+\delta)$) for different values of applied current $I$. The width of
the simulated ladder is $w=40\xi(0)$ and the size of the holes is $a=20\xi(0)$. The temperature is $T=0.98T_c$ and the
GL parameter is $\kappa=10$.}
\end{figure}

As a representative example, we studied the response to an external magnetic field of a superconducting strip of width
$w=40\xi(0)$ and with holes of size $a=20\xi(0)$. Figure \ref{fig2}(a) shows the time averaged energy $F$ (see Ref. \cite{swei} for the definition of $F$) of this
system as a function of magnetic field for different values of the applied dc current $I$. For small current values
(curve 1), the vortex entries are shifted to lower magnetic field compared to the case of no applied current (not shown
here). Still, vortex entry remains a first-order transition, i.e., one observes a sharp drop in the energy at the
transition field (indicated by a solid black arrow in Fig. \ref{fig2}(a)). However, for the same current at larger
magnetic fields we surprisingly found that vortex entry is characterized by a {\it continuous change in energy},
manifested by a second-order transition (dashed blue arrow). It is exactly these transitions that give rise to the
large voltage signal $V=-\int\partial {\bf A}/\partial t\cdot {\bf dl}$, as shown in Fig. \ref{fig2}(b) (see curve 1). Such vortex entries are found at lower fields for
larger applied current (curves 2 in Fig. \ref{fig2}(a)). With further increasing $I$, the magnetic fields at which
voltage oscillations are located do not change, but oscillations become broader (curves 3 and 4 in Fig. \ref{fig2}(b)).
Note that oscillations in our curves are not exactly periodic with field and larger flux is needed to observe the first
voltage peak (a typical effect for finite-walled ring structures, see e.g. Ref. \cite{loop2}).

\begin{figure}[b]
\includegraphics[width=\linewidth]{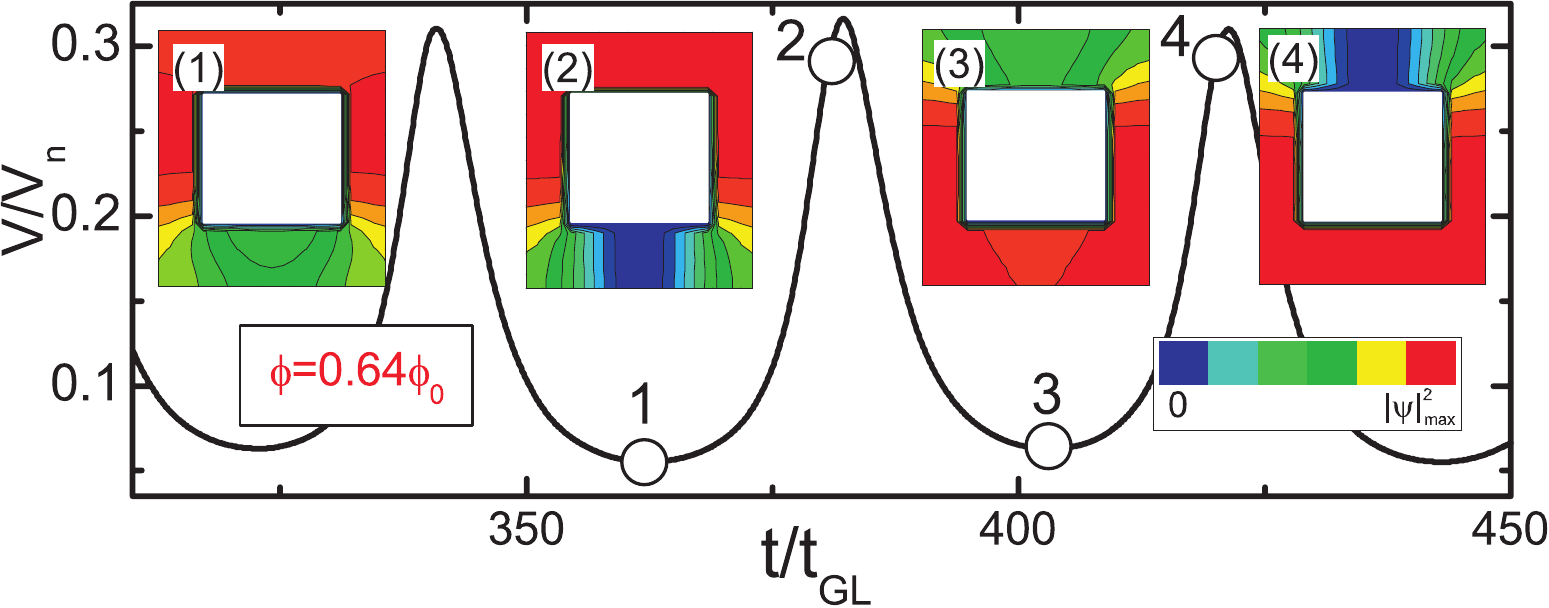}
\caption{\label{fig3}(color online) Voltage vs. time response for the sample of Fig. \ref{fig2} for $\phi=0.64\phi_0$ and $I/(j_0w)=0.04$. Insets show snapshots of $|\psi|^2$ at time intervals indicated by open circles on the $V(t)$ curves.}
\end{figure}

\begin{figure}[t]
\includegraphics[width=\linewidth]{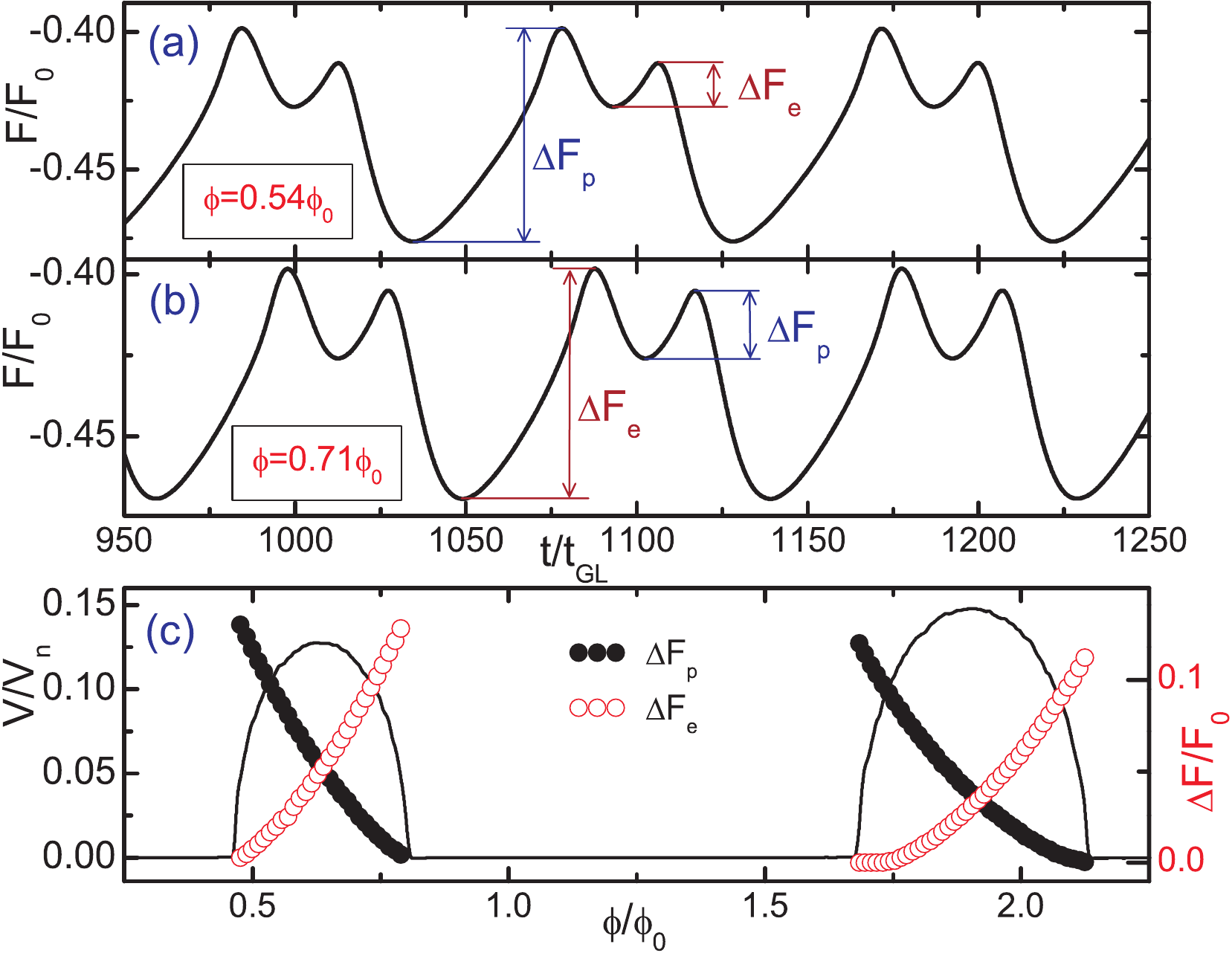}
\caption{\label{fig4}(color online) (a,b) Time evolution of the free energy for the ladder considered in Fig.
\ref{fig2} at $\phi=0.54\phi_0$ (a) and $\phi=0.71\phi_0$ (b). Arrows indicate the penetration $\Delta F_p$ and
expulsion $\Delta F_e$ barriers. (c) Time-averaged voltage (solid curve), $\Delta F_p$ (filled circles) and $\Delta
F_e$ (open circles) as a function of external flux. The applied current is $I/(j_0w)=0.04$.}
\end{figure}

To gain further insight into the process leading to magnetoresistance peaks, we plot the temporal voltage signal for
$I/(j_0w)=0.04$ and $\phi=0.64\phi_0$ in Fig. \ref{fig3}. At this field (i.e., just after the maximum of the first
voltage peak), the output voltage oscillates periodically in time with a global minimum (point 1 in Fig. \ref{fig3})
corresponding to one vortex trapped inside the hole (inset 1). With time, this vortex exits the sample (inset 2)
exhibiting a maximum in the $V(t)$ characteristics (point 2), and the system is free of vortices (inset 3). Immediately
after, a new vortex penetrates the sample (inset 4) and the entire vortex entry-exit sequence repeats (see Ref.
\cite{suppo} for an animated data for this process).

We stress once more that vortices in our study cannot be induced by thermal fluctuations, making our predictions
different from ones in Refs. \cite{Kirtley, Sochnikov}. Note also that no voltage/resistance oscillations with time are
expected from the LP effect, which is also different from our findings. The origin of vortex excitation in our system
is the {\it interplay of currents}. Namely, applied magnetic field induces reactive screening (and circulating)
currents in the superconductor, $j_s$. Therefore, applied dc current always enhances the supercurrent $j_s$ on one side
of the sample, while suppressing $j_s$ on the other side. This directly affects the barriers for vortex entry and exit,
which are indicated in Figs. \ref{fig4}(a,b), where we plotted the time evolution of the free energy of the system.
When the total current $j_t=j_s+j$ reaches to its critical value \cite{denis_barrier} in one arm of the ladder, a
vortex nucleates at the edge of the sample in spite of the finite energy barrier $\Delta F_p$ (see filled circles in
Fig. \ref{fig4}(c)). At that field the barrier for the vortex expulsion $\Delta F_e$ on the opposite side of the sample
is strongly suppressed (see open circles in Fig. \ref{fig4}(c)), since $j_t=j_s-j$. Therefore, the vortex is driven
through the system resulting in periodic-in-time voltage signal as shown in Fig. \ref{fig3}. With further increasing
the applied field, the barrier for vortex exit increases (open circles in Fig. \ref{fig4}(c)). Consequently, the vortex
remains trapped in the sample and voltage becomes zero. Such modulation of the barrier for vortex entry and exit is
observed for the transition between different vortex states (see the second voltage peak in Fig. \ref{fig4}(c)).

Thus, we conclude that each voltage (and magnetoresistance) peaks as a function of magnetic field corresponds to the
moving vortices in the system, the motion of which is manipulated by the interplay of applied and Meissner currents.
Note however that at higher driving current or for larger vorticity, it is no longer possible to stabilize the
stationary vortex state and vortices always remain in motion, creating a background voltage (characteristic of a
resistive state) on which the peaks for every additional vortex entry are superimposed (see curve 4 in Fig.
\ref{fig2}(b)). For this case, animated data for the evolution of the superconducting condensate with time, together
with the corresponding voltage signal, are shown in the Supplementary Material \cite{suppo}. The background on which
the magnetoresistance oscillations are superimposed has been puzzling scientists since the original experiment of
Little and Parks \cite{Little}. Several possible explanations have been put forth, including the Meissner effect,
misalignment of the external field, and different single-particle excitation energies within an electron pair. Recently
Sochnikov {\it et al.}, proposed an alternative mechanism which is based on thermally excited vortices and antivortices
\cite{Sochnikov}. In our case, the background stems solely from the continuous, dissipative vortex motion.

\begin{figure}[t]
\includegraphics[width=\linewidth]{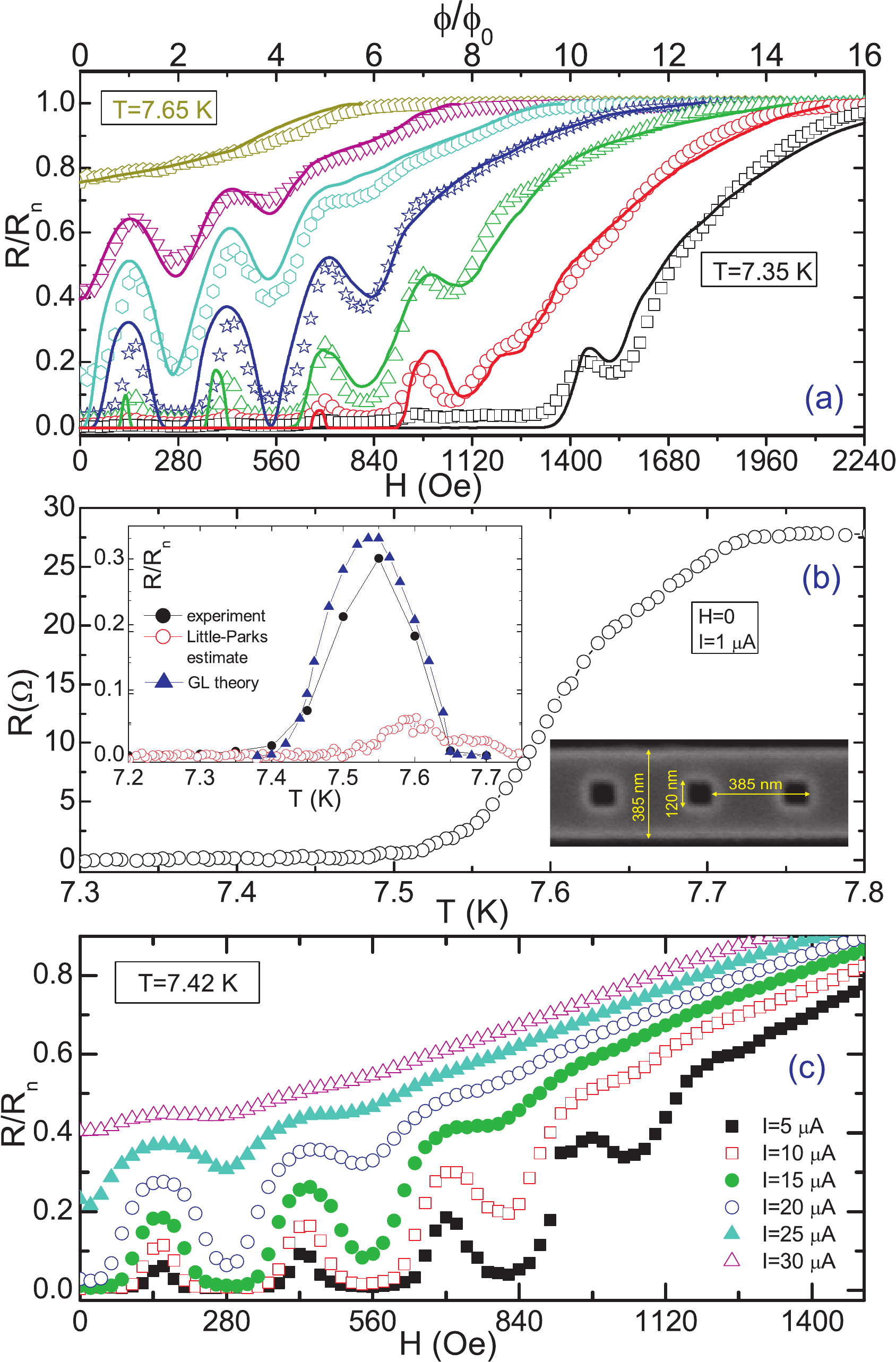}
\caption{\label{fig5}(color online) (a) Measured resistance (in units of normal state resistance $R_n=27.9~\Omega$) vs.
magnetic field for a Nb superconducting strip (of width $w=385$ nm) with a chain of holes (of size $a=120$ nm and
period $d=385$ nm), in applied dc current $I=1~\mu$A, and at different temperatures (ranging from $T=7.35$ K to
$T=7.65$ K with step 50 mK). Solid curves show theoretical results for the dimensions of the experimental sample and
for $I/(j_0w)=0.05$. (b) $R(T)$ curve of the sample at zero magnetic field for $I=1~\mu$A. Lower inset shows the SEM
image of the sample. Upper inset shows the amplitude of the first resistance peak, $\Delta R$, as a function of
temperature (solid dots), compared to the amplitude of the resistance oscillations predicted by the Little-Parks
effect, $\Delta R=0.14(dR/dT)T_c(\xi(0)/(w+a)/4)^2$ \cite{Little,Sochnikov} (open dots). Triangles present the results
of the GL simulations. (c) $R(H)$ curves of the sample obtained at $T=7.42$ K for different values of the bias
current.}
\end{figure}

\textit{Experiment}. -- In order to confirm our theoretical prediction, we performed magnetoresistance measurements on
100 nm thick Nb strips with a chain of square holes patterned with focused-ion-beam milling (see Ref. \cite{hua} for
the details of sample fabrication and measurements). We considered Nb strips with width $w$=385 nm, hole size $a$=120
nm and period $d$=300 nm, 385 nm and 800 nm. Magnetoresistance of the samples was obtained by four-probe dc
measurements using a constant current mode. Here we present the results for the sample with interhole spacing $d$=385
nm (see the lower inset in Fig. \ref{fig5}(b)), which shows a broad superconducting/normal transition with
$T_c\approx7.725$ K (using the criterion of 90 \% of the normal state resistance, see Fig. \ref{fig5}(b)). Zero
temperature coherence length $\xi(0)$ and the penetration depth $\lambda(0)$ were estimated to be 10 nm and 200 nm,
respectively.

Figure \ref{fig5}(a) shows experimentally measured magnetoresistance curves of the sample at different temperatures
(see Supplemental Material \cite{sup300nm} for the results obtained for the sample with $d$=300 nm). Because of the
small bias current ($I=1~\mu$A and current density $j=2.5\cdot10^3$ A$\cdot$cm$^{-2}$), no resistance oscillations are
observed until higher temperatures ($T=7.3$ K), where a first resistance peak appears at around 1400 Oe. Figure
\ref{fig5}(c) shows the $R(H)$ curves obtained at $T=7.42$ K for different values of the biased current. As delineated
in this figure, the amplitude of the resistance oscillations is strongly affected by the bias current: the oscillations
disappear both at larger and smaller currents. However, the position of the resistance peaks is independent of the
applied current, as we predicted in our numerical simulations (see Fig. \ref{fig2}(b)).

We now analyze the amplitude of the resistance oscillations as a function of temperature, shown in the top inset of
Fig. \ref{fig5}(b) by solid dots for the first peak of the $R(H)$ curves of Fig. \ref{fig5}(a). Open dots in this
figure show the expected amplitude of the LP oscillations due to periodic changes in the superconducting transition
temperature: $\Delta R=0.14(dR/dT)T_c(\xi(0)/r_{eff})^2$ with $r_{eff}=(w+a)/4$ \cite{Little,Sochnikov}. The LP
mechanism predicts resistance oscillations at higher temperatures, $T>7.5$ K, and can survive until very close to
$T_c$. However, we observed resistance peaks already starting at $T=7.35$ K, and which disappear well below $T_c$.
Moreover, the amplitude of the oscillations reported here is much larger than the ones originating from the LP effect,
in spite of the fact that the LP effect is pronounced in low-$T_c$ superconductors due to their relatively large
coherence length. On the other hand, current-driven moving vortices can account for the observed large
magnetoresistance. The solid curves in Fig. \ref{fig5}(a) are results of theoretical simulations for the parameters of
the experimental sample. We found very good agreement with the period of the voltage oscillations, while we observed a
weak difference in the amplitude of the oscillations - likely due to the presence of superconducting current and
voltage leads in the experiment (see triangles in the top inset of Fig. \ref{fig5}(b)). In simulations, the damage to
the sample near the (hole) edges due to the ion bombardment process (the bright regions in the SEM image of the sample
at the sample edges) was modeled by the local suppression of $T_c$  of approx. 4 \% in these regions ($\sim$ 20 nm
width). We also used the parameter $u=1$ (see Eq. (1)) to best fit the experimental data, which indicates a larger
electric field penetration depth $l_E$ in our experimental samples (since $u\sim l_E^{-2}$) \cite{Ivlev}. The
interpretation of the voltage oscillations vs. magnetic field remains the same as in Figs. \ref{fig2}-\ref{fig4},
confirming their origin to be the moving vortices. Here we must emphasize that due to the small width of the rings
(i.e., $(w-a)/2\lesssim 3\xi(T)$) in the considered temperature range, the model of thermally created vortex-antivortex
pairs (Ref. \cite{Sochnikov} and references therein) is not applicable to our experimental sample.

In summary, we have demonstrated that the Little-Parks effect, a
landmark manifestation of flux quantization, is not sufficient to
fully describe the size and temperature range of typical
magnetoresistance oscillations observed in small superconductors.
Viable interpretations of such oscillations have been offered in the
past by Parks and Mochel \cite{parks}, Anderson and Dayem
\cite{anderson}, Sochnikov {\it et al.} \cite{Sochnikov} and
Baturina {\it et al.} \cite{baturina}. We however showed that in
low-T$_c$ mesoscopic samples large magneto-resistance oscillations
can stem solely from intermittent vortex nucleation, motion, and
stabilization by the interplay of the driving current and the
persistent Meissner currents in the sample. Furthermore,
continuously moving (dissipative) vortices contribute to the
background of the magnetoresistance curves, and their contribution
grows with increasing field or applied current - an effect not
covered by competing explanations. We therefore suggest that, due to
ever present current in transport measurements, the contribution of
current-induced vortices to dissipation in patterned superconductors
can never be entirely disregarded.

This work was supported by the Flemish Science Foundation (FWO-Vl) and the Belgian Science Policy (IAP) (theory) and by
the US Department of Energy (DOE) Grant No. DE-FG02-06ER46334 (experiment). G.R.B. acknowledges individual grant from
FWO-Vl. W.K.K. acknowledges support from DOE BES under Contract No. DE-AC02-06CH11357, which also funds Argonne's
Center for Nanoscale Materials (CNM) where the focused-ion-beam milling was performed. M.L.L was a recipient of the
NIU/ANL Distinguished Graduate Fellowship.

\end{document}